\documentclass[aps,pra,twocolumn,superscriptaddress,nofootinbib,longbibliography]{revtex4-2}

\usepackage[T1]{fontenc}
\usepackage[utf8]{inputenc}
\usepackage{amsmath,amssymb,bm}
\usepackage{braket}
\usepackage{xcolor}
\usepackage{graphicx}
\usepackage{hyperref}
\usepackage{orcidlink}

\newcommand{\FusionRO}{\ensuremath{\mathrm{FR}}}
\newcommand{\PSQWC}{\ensuremath{\mathrm{PS}_{\mathrm{QWC}}}}
\newcommand{\MSE}{\ensuremath{\mathrm{MSE}}}

\graphicspath{{Figures/}}

\begin{document}

\title{Native topological readout on qubit hardware: a Fibonacci-chain benchmark of measurement-compilation trade-offs}

\author{Babatunde Moses Ayeni\texorpdfstring{\,\orcidlink{0000-0003-4035-149X}}{}}
\affiliation{Department of Physics, Maynooth University}
\affiliation{School of Mathematics and Statistics, Technological University Dublin}

\date{\today}

\begin{abstract}

Recent demonstrations of non-Abelian braiding of graph vertices on noisy intermediate-scale quantum (NISQ) superconducting processor, and the experimental realization of topological order in general on various quantum hardware platforms necessitate an important question: when does a native (topological) fusion readout genuinely help for topological anyonic Hamiltonians implemented on NISQ hardware? We use the Fibonacci anyons chain as a concrete model for understanding the trade-off between measurement cost and compilation cost in that setting. The comparison is made against a simple grouped-Pauli baseline, and is scored by a covariance-aware mean-squared-error (MSE) of the full energy estimator. We based our benchmark on two different important classes of quantum circuits, namely Floquet time-evolved and variational quantum eigensolver quantum circuits, with the underlying Hamiltonian consisting of both braiding and fusion interaction. Our analysis found that there is not a uniform best method across both problems: the fusion readout method performed better on Floquet-type circuits on both the MSE and covariance-aware sampling variance, while the grouped Pauli method performed better on VQE on the MSE but worse on sampling variance. We derive scaling laws, and compute shot-budget crossover points, where one method is operationally favored above the other. The relevance of this work extends beyond Fibonacci chains to two-dimensional topological models compiled on superconducting and other qubit-native platforms, and can be used as a guide in answering the question of when one should measure in the native operator basis of the target physics, or when it is better to fall back on Pauli-basis reconstruction. 
\end{abstract}

\maketitle

\section{Introduction}

Near-term quantum algorithms are constrained not only by coherent gate error but also by measurement cost: the number of circuits, shot pools, and classical post-processing steps required to estimate observables to useful precision~\cite{preskill2018quantum,peruzzo2014variational,mcclean2016theory}. This becomes particularly severe when a target Hamiltonian is hosted on qubit hardware but its physically natural observables are not short Pauli sums in the device basis. An important class of such Hamiltonians are topological models with non-Abelian anyonic excitations. Operationally, the issue is simple: a qubit processor only returns qubit readout data, so an encoded topological observable must be accessed either by reconstructing it from Pauli-basis measurements or by compiling an extra unitary that rotates the state into a basis where ordinary qubit readout implements the desired projector or fusion measurement. A large literature has therefore focused on making Pauli-frame estimation more efficient through commuting-group strategies, adaptive allocation, low-rank factorizations, and shadow-like methods~\cite{verteletskyi2020measurement,Gokhale2019,gokhale2020optimization,Izmaylov2019,kuebler2020adaptive,gu2021adaptive,huggins2021efficient,huang2020predicting}. Recent refinements also include exhaustive searches for larger commuting measurement groupings within VQE-style estimators~\cite{YenFullCommuting2023}. Those advances sharpen the question that matters for structured models: when should one remain in the Pauli frame and optimize within it, and when is it better to leave that frame and pay additional compilation cost to measure in a basis adapted to the target physics?

We use the Fibonacci anyons chain with both topological braiding and fusion interaction as a sort of minimal model to study this problem. That question is broader than the Fibonacci chain alone. It arises whenever topological or projector-dominated models are compiled onto qubit-native processors: in anyonic topological engineering on superconducting hardware, in string-net or Levin-Wen-type Hamiltonians represented on superconducting quantum processors, and in code-space or defect-based braiding architectures, the target observables are often most natural in fusion, projector, or constraint variables rather than in the bare Pauli language of the host device~\cite{levin2005stringnet,gottesman2002introduction,dennis2002topological,fowler2012surface,bombin2010topological,brown2017poking,bernien2017probing,browaeys2020manybody}. Recent processor-level demonstrations now make that broader motivation more concrete, including non-Abelian braiding of graph vertices on superconducting hardware, trapped-ion demonstrations of non-Abelian topological order, and superconducting realizations targeted specifically at Fibonacci anyons.~\cite{Google2023,Quantinuum2024,XuFibonacci2024} Once such models are hosted on qubit hardware, one still faces a measurement-design choice: reconstruct the observable in the Pauli frame of the processor, or compile a measurement into the native operator basis of the encoded topological model.

Fibonacci anyon chains provide a natural flagship setting for studying that trade-off. Their local observables are fusion-channel projectors, and the corresponding native basis changes are written directly in terms of the same $F$- and $R$-matrices primitives that define the theory~\cite{Field2018,nayak2008nonabelian,pachos2012introduction,pfeifer2015finite,singh2014matrix,ayeni2016simulation,kirchner2023numerical}. This is why fusion readout is more than an arbitrary alternative measurement prescription: it is a readout strategy aligned with the observable structure of the Hamiltonian itself. 

It is important to state that we model Fibonacci anyons on quantum hardware through an encoding of the fusion space, and not through a direct engineering of anyonic excitations, as in some recent experimental papers.\cite{Google2023,Quantinuum2024,XuFibonacci2024} Still the question we asked remains relevant once topological models are compiled onto non-native quantum processors (e.g. superconducting): if the target observable is native in the encoded anyonic description, should it still be measured through Pauli reconstruction, or should one pay the extra basis-change depth required to measure it more natively? \footnote{One can think of the encoded problem as the mathematical model of the physical anyons topological engineering problem. So, the measurement-compilation cost tradeoff in the physical problem should similarly manifest in the simulation of its mathematical model.}

We organize our benchmark around a covariance-aware fixed-budget \MSE\ of the full energy estimator, and not around covariance in isolation. The reference baseline chosen is the grouped Pauli sampling with qubit-wise-commuting groups (\PSQWC), because it is transparent and interpretable rather than because it is the strongest possible generic Pauli estimator. The aim is this work is not to find a single winner in the measurement-optimization problem, but to test whether a measurement rule that is physically aligned with a topological Hamiltonian can retain its estimator-level advantage after compilation to a constrained NISQ backend.

The paper makes four concrete points. First, it formulates the estimator-level criterion that tests whether a native measurement advantage survives at fixed shot budget. Second, it benchmarks the fusion readout (\FusionRO) against the qubit-wise grouped Pauli strategy (\PSQWC) on a digital Floquet-type evolution, and on optimized VQE shallow circuits of Fibonacci-chain Hamiltonians. Third, it shows that noiseless and hardware comparisons need not agree: \FusionRO\ wins uniformly on covariance-aware sampling variance, but not uniformly on realized estimator error. Fourth, it connects that crossover to a resource-proxy analysis derived from the same scaling-law fits used in the main results. The upshot of our analysis can be stated simply as, native fusion readout provides a real sampling advantage, but compiled hardware cost can still determine the practical winner under NISQ constraints. Even if this could have been surmised intuitively, our work demonstrates this rigorously both qualitatively and quantitatively. 

The rest of this article is organized as follows. Section~II defines the Fibonacci-chain Hamiltonian and explains why fusion readout is native to its local observables. Section~III develops the covariance-aware fixed-budget estimator framework and states the criterion used to compare \FusionRO\ with grouped Pauli reconstruction. Section~IV specifies the two measurement strategies, the digital Floquet and optimized-state VQE workloads, and the matched-budget hardware protocol. Section~V presents the noiseless and hardware results, Section~VI discusses their implications for compiled topological measurements on NISQ hardware, and Section~VII concludes. The appendices collect the Fibonacci $F$/$R$ conventions, selected crossover-scaling grids, the resource-proxy scaling analysis, and the hardware transpilation methodology.

\section{Fibonacci Anyons Chain Model and Native Observables}

We consider one-dimensional Fibonacci anyon chains in the fusion-path encoding~\cite{pfeifer2015finite,singh2014matrix,ayeni2016simulation,ayeni2018,kirchner2023numerical}. In this encoding, each qubit stores an intermediate fusion label, with $\ket{0}\leftrightarrow \mathbf{1}$ and $\ket{1}\leftrightarrow \tau$, and only fusion-consistent bit strings represent physical states. The local observables are short-range projector-type terms, making the model a natural setting for comparing Pauli-frame estimation with a native anyonic readout.

The Hamiltonian studied throughout this work consists of neighbouring and next-neighbouring interactions, which can be written as
\begin{equation}
H_n = J_{\mathrm{F3}} \sum_{i=0}^{n-3} H_i^{\mathrm{F3}} + J_{\mathrm{BF4}} \sum_{i=0}^{n-4} H_i^{\mathrm{BF4}},
\label{eq:chain-hamiltonian}
\end{equation}
with local terms
\begin{align}
H_i^{\mathrm{F3}} &= F_i^{\dagger} Z_{i+1} F_i, \\
H_i^{\mathrm{BF4}} &= B_i^{\dagger} F_{i+1}^{\dagger} Z_{i+2} F_{i+1} B_i,
\label{eq:local-terms}
\end{align}
where $H_i^{\mathrm{F3}}$ is the $(\tau_i,\tau_{i+1})$ neighbouring interaction term that maps to three qubits and $H_i^{\mathrm{BF4}}$ is the $(\tau_i, \cdot, \tau_{i+2})$ next-neighbouring interaction term that maps to four qubits, as shown below.  Here $F_i$ is the local recoupling move on the (fusion tree) window $(i,i{+}1,i{+}2)$ and $B_i$ is the braid operator in the same fusion-tree language, where $B_i$ might be composed of the F- and R-matrices as appropriate. Throughout our benchmark we fix the values of the coefficients as $J_{\mathrm{F3}}=1.0$ and $J_{\mathrm{BF4}}=0.5$.
Appendix~\ref{app:fib-background} records the explicit Fibonacci $F$,$R$, and $B$ conventions and the implementation-level braid identity used by the compiled circuits. Therefore, in the fusion-path encoding, the entire Hamiltonian of $n$ anyons maps onto $n+1$ qubits.

The structural reason that fusion readout is native can be seen easily by considering the BF4 term: $H_i^{\mathrm{BF4}}$, which represents the interaction between $\tau_i$ and $\tau_{i+2}$. To foster this interaction, both anyons have to be brought into neighbouring positions using a braid move $B_i$. Let $h_{i+1}^{\mathrm{DF}}$ now be the local (projector-type) observable acting on the two neighbouring anyons $(\tau_{i+1}, \tau_{i+2})$ in a direct fusion. The physical interaction may be completely written as
\begin{equation}
H_i^{\mathrm{BF4}} = B_i^{\dagger} F_{i+1}^{\dagger} h_{i+1}^{\mathrm{DF}} F_{i+1} B_i.
\label{eq:bf4-conjugation}
\end{equation}
The details are contained in the Appendix. For an arbitrary prepared state $\rho$, the corresponding local expectation value admits two equivalent trace forms
\begin{align}
\langle H_i^{\mathrm{BF4}}\rangle_{\rho}
&= \mathrm{Tr}\!\left(B_i^{\dagger}F_{i+1}^{\dagger} h_{i+1}^{\mathrm{DF}} F_{i+1} B_i\,\rho\right) \label{eq:bf4-traces_1} \\
&\equiv \mathrm{Tr}\!\left(h_{i+1}^{\mathrm{DF}}\,F_{i+1} B_i\,\rho\,B_i^{\dagger} F_{i+1}^{\dagger}\right).
\label{eq:bf4-traces}
\end{align}
Equation~\eqref{eq:bf4-traces_1} and ~\eqref{eq:bf4-traces} already contains the two measurement logics used in the paper. In \PSQWC, one measures a grouped Pauli decomposition of the mapped operator content on the state in Eq.~\eqref{eq:bf4-traces_1}. In \FusionRO, one instead appends the basis change $U_i = F_{i+1}B_i$ to the state-preparation circuit and measures the direct-fusion observable $h_{i+1}^{\mathrm{DF}}$ directly on the rotated state as in Eq.~\eqref{eq:bf4-traces}. The same logic extends termwise to the full Hamiltonian in Eq.~\eqref{eq:chain-hamiltonian}. Even though both methods are equivalent in principle, in practice they behave differently on NISQ processor, where they have different compilation and measurement-error cost.

\begin{figure}[t]
\centering
\includegraphics[width=0.96\linewidth]{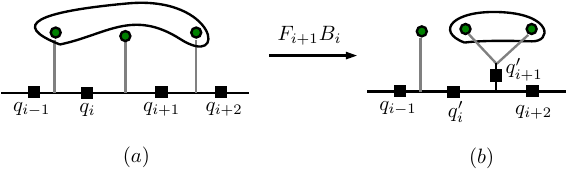}
\caption{Native-measurement motivation for the Fibonacci-chain Hamiltonian. The local basis change $U_i = F_{i+1}B_i$ maps the BF4 interaction into a direct-fusion frame where the corresponding observable is a simple local projector. Fusion readout is therefore aligned with the observable structure of the model rather than introduced as an arbitrary competitor in a generic measurement comparison.}
\label{fig:native-schematic}
\end{figure}

The next section turns this structural observation into an estimator-level comparison framework. It defines the fixed-budget energy estimator, separates covariance-driven sampling effects from hardware-sensitive contributions, and derives the crossover criterion used to decide when the native sampling advantage of \FusionRO\ survives compiled execution.

\section{Estimator Framework and Native-Measurement Criterion}
\label{Sec:Measurement criterion} 

For the purposes of the benchmark, the Hamiltonian is written abstractly as
\begin{equation}
H = \sum_{\ell=1}^{L} w_{\ell} O_{\ell},
\label{eq:general-hamiltonian}
\end{equation}
where the $O_{\ell}$ are fixed local observables and $w_{\ell}$ the corresponding weights, which we write as a vector $\bm{w} = (w_1, \ldots, w_L)^{\mathrm{T}}$. For each method $m\in\{\FusionRO,\PSQWC\}$ and total shot budget $N$, the full-Hamiltonian estimator is
\begin{equation}
\hat E_m^{(N)} = \sum_{\ell=1}^{L} w_{\ell} \hat o_{\ell,m}^{(N)} = \bm{w}^{\mathsf T} \hat{\bm{o}}_m^{(N)},
\label{eq:full-estimator}
\end{equation}
where $\hat{\bm{o}}_m^{(N)}$ is a vector of the estimators of each fixed local observable $O_{\ell}$.

The benchmark quantity is the fixed-budget mean-squared error
\begin{equation}
\MSE_m(N) = \mathbb{E}\!\left[\left(\hat E_m^{(N)}-E_{\mathrm{exact}}\right)^2\right].
\label{eq:mse-def}
\end{equation}
Writing the mean estimator value as
\begin{equation}
\mu_m(N) = \mathbb{E}\!\left[\hat E_m^{(N)}\right],
\end{equation}
and the bias term as
\begin{equation}
b_m(N)=\mu_m(N)-E_{\mathrm{exact}},
\end{equation}
the standard add-and-subtract step gives
\begin{equation}
\MSE_m(N) = b_m(N)^2 + \mathrm{Var}\!\left(\hat E_m^{(N)}\right).
\label{eq:bias-var}
\end{equation}
This is the quantity we used to benchmark the cost of doing measurement in the native \FusionRO\ method against the simple \PSQWC\ method. 

%that matters for the comparison. Any shot-level derivation is only pedagogical unless it is reassembled into the averaged estimator in Eq.~\eqref{eq:full-estimator}.
%Equivalently, one may imagine a scalar per-shot route in which all method-specific reconstruction has been absorbed into random contributions $X_{k,m}$ satisfying $\hat E_m^{(N)}=N^{-1}\sum_{k=1}^{N} X_{k,m}$. In that language, $\mathbb{E}[(X_{k,m}-E_{\mathrm{exact}})^2]$ is a per-shot error quantity rather than the benchmark target itself. The correct add-and-subtract step must therefore use the fixed mean $\mu_m(N)=\mathbb{E}[\hat E_m^{(N)}]$, not the random estimator on the right-hand side of the same identity, and the covariance-aware term reappears only after reconstructing the averaged estimator in Eq.~\eqref{eq:full-estimator}.

To isolate sampling variance from hardware noise, we split the estimator variance as 
\begin{equation}
\mathrm{Var}\!\left(\hat E_m^{(N)}\right)
=
\mathrm{Var}_{\mathrm{sample},m}(N)
+
\mathrm{Var}_{\mathrm{hw},m}(N),
\label{eq:var-split}
\end{equation}
where the covariance-aware sampling term is
\begin{align}
\mathrm{Var}_{\mathrm{sample},m}(N)
&=
\bm{w}^{\mathsf T}\Sigma_m(N)\bm{w}, \\
\Sigma_m(N)
&=
\mathrm{Cov}\!\left(\hat{\bm{o}}_m^{(N)}\right),
\label{eq:sampling-var}
\end{align}
and $\mathrm{Var}_{\mathrm{hw},m}(N)$ is the variance term due to hardware noise.
At fixed circuit family and fixed shot-allocation fractions we write the large-$N$ form as
\begin{equation}
\Sigma_m(N)=\frac{\widetilde\Sigma_m}{N}+o\!\left(N^{-1}\right),
\qquad
\Gamma_m = \bm{w}^{\mathsf T}\widetilde\Sigma_m\bm{w},
\end{equation}
so that the leading sampling variance contribution is $\Gamma_m/N$.

This motivates the leading-order model
\begin{equation}
\MSE_m(N) \approx b_m(N)^2 + \frac{\Gamma_m + \kappa_m}{N},
\label{eq:leading-mse}
\end{equation}
where $\kappa_m$ absorbs the hardware-sensitive $1/N$ contribution. The difference between methods is therefore controlled by
\begin{align}
\Delta_{\MSE}(N)
&=
\MSE_{\FusionRO}(N)-\MSE_{\PSQWC}(N) \\
&\approx
\Delta b^2 + \frac{\Delta\Gamma_{\mathrm{eff}}}{N},
\label{eq:delta-mse}
\end{align}
with $\Delta b^2 = b_{\FusionRO}^2-b_{\PSQWC}^2$ and $\Delta\Gamma_{\mathrm{eff}} = (\Gamma_{\FusionRO}+\kappa_{\FusionRO})-(\Gamma_{\PSQWC}+\kappa_{\PSQWC})$. The crossover point (where $\Delta_{\mathrm{MSE}}(N) = 0) $ is obtained as $N_c \approx - \Delta\Gamma_{\mathrm{eff}}/\Delta b^2$.
If we make the reasonable assumptive choice that $\Delta\Gamma_{\mathrm{eff}}<0$ and $\Delta b^2 > 0$, as supported empirically by data below, then the characteristic crossover scale is
\begin{equation}
N_c \approx \frac{|\Delta\Gamma_{\mathrm{eff}}|}{\Delta b^2}.
\label{eq:crossover}
\end{equation}
This is the central criterion of the paper: native fusion readout is useful only when its covariance advantage survives the non-sampling penalty introduced by compiled hardware execution.

\section{Measurement Strategies and Benchmark Workloads}

In this section we describe the measurement strategies and the quantum problems. We first specify how \FusionRO\ and \PSQWC\ estimate the same local Hamiltonian terms, then define the two quantum circuit families used throughout the comparison: digital Floquet-evolved states and optimized-state VQE circuits. We close by stating the matched shot-budget rule and the hardware-execution protocol that put both measurement strategies on the same footing.

\subsection{Fusion readout and grouped Pauli reconstruction}

For fusion readout, each local term is measured in a problem-adapted native frame. Concretely, for each observable $O_{\ell}$ we compile a short basis change $U_{\ell}$ such that
\begin{equation}
O_{\ell}=U_{\ell}^{\dagger} h_{\ell}^{\mathrm{native}} U_{\ell},
\end{equation}
with $h_{\ell}^{\mathrm{native}}$ diagonal or otherwise low support in the encoded qubit basis. The local estimator is then
\begin{align}
o_{\ell}
&= \mathrm{Tr}\!\left(\rho O_{\ell}\right)
= \mathrm{Tr}\!\left(U_{\ell}\rho U_{\ell}^{\dagger} h_{\ell}^{\mathrm{native}}\right), \\
\hat o_{\ell,\FusionRO}^{(r)}
&= \frac{1}{N_{\ell,r}}\sum_{s=1}^{N_{\ell,r}} x_{\ell,s}^{(r)},
\qquad
x_{\ell,s}^{(r)}\in\mathrm{spec}\!\left(h_{\ell}^{\mathrm{native}}\right),
\label{eq:fr-local-estimator}
\end{align}
where $r$ labels replicates. For the Fibonacci-chain Hamiltonian, the number of native measurement circuits is one per local term, hence $2n-5$ circuits at qubit number $n$.

For \PSQWC, the same local observable is reconstructed from a grouped Pauli expansion,
\begin{align}
O_{\ell} &= \sum_{\alpha=1}^{M} a_{\ell\alpha} P_{\alpha},
\qquad
\{1,\ldots,M\}=\bigsqcup_{g=1}^{G} \mathcal{G}_g, \\
\hat o_{\ell,\PSQWC}^{(r)}
&= \sum_{\alpha=1}^{M} a_{\ell\alpha}\,\hat p_{\alpha}^{(r)},
\label{eq:ps-local-estimator}
\end{align}
where each group $\mathcal{G}_g$ is qubit-wise commuting and the reconstructed Pauli expectations $\hat p_{\alpha}^{(r)}$ are extracted from the shared group counts. 

The hierarchy of sampled objects matters. One shot is one execution of one measurement circuit. One replicate is one full rerun of the complete estimator at fixed total budget $N$, including all method-specific sub-allocations. The estimator is therefore fundamentally defined over shots, while replicates are used only to estimate fixed-budget moments of that estimator empirically.

Collecting the local estimators into $\hat{\bm{o}}_m^{(r)}$ for replicate $r$, the Hamiltonian-energy estimate and the covariance-aware sampling-variance estimator are
\begin{equation}
\hat E_m^{(r)} = \bm{w}^{\mathsf T}\hat{\bm{o}}_m^{(r)},
\qquad
\hat{\mathrm{Var}}_m^{(r)} = \bm{w}^{\mathsf T}\Sigma_m^{(r)}\bm{w}.
\label{eq:energy-and-var-est}
\end{equation}
Across $R$ replicates, the empirical fixed-budget estimators for the energy and MSE are
\begin{align}
\bar E_m &= \frac{1}{R}\sum_{r=1}^{R} \hat E_m^{(r)}, \\
\widehat{\MSE}^{\mathrm{emp}}_m
&= \frac{1}{R}\sum_{r=1}^{R}\left(\hat E_m^{(r)}-E_{\mathrm{exact}}\right)^2.
\label{eq:empirical-mse}
\end{align}

For the empirical \MSE\ itself, we quantify the within-cell uncertainty by the replicate standard error of the squared errors,
\begin{equation}
\begin{split}
\widehat{\mathrm{SE}}_{\MSE,m}
&=
\frac{s_{\mathrm{sq},m}}{\sqrt{R}}, \\
s_{\mathrm{sq},m}^2
&=
\frac{1}{R-1}\sum_{r=1}^{R}
\left[
\left(\hat E_m^{(r)}-E_{\mathrm{exact}}\right)^2
-\widehat{\MSE}^{\mathrm{emp}}_m
\right]^2,
\end{split}
\label{eq:mse-se}
\end{equation}
and use the corresponding pooled FR-versus-PS standard error below as a descriptive uncertainty proxy for cellwise empirical-\MSE\ differences.

\subsection{Digital Floquet evolution}

The digital benchmark uses a Floquet-type time evolution to generate states that can be used as a controlled family of comparison data without a classical optimizer feedback. 

With the Hamiltonian from Eq.~\eqref{eq:chain-hamiltonian}, the logical product-formula circuit is
\begin{equation}
V(s)=\Bigl(\prod_j e^{-i\,\delta t\,H_j}\Bigr)^s,
\qquad
\rho_n(s)=V(s)\rho_0 V^{\dagger}(s),
\label{eq:trotter}
\end{equation}
where $s$ is the Trotter-step count and $\rho_0=\ket{1\cdots 1}\!\bra{1\cdots 1}$ is the fixed fusion-valid reference state used throughout the benchmark. One step is built from ordered sweeps over the F3 and BF4 terms,
\begin{align}
U_{\mathrm{A}} &= \prod_{j=0}^{n-3}\Bigl(U_F^{(j)} e^{-i\,\delta t\,J_{\mathrm{F3}}\, Z_{j+1}} U_F^{(j)\dagger}\Bigr), \\
U_{\mathrm{B}} &= \prod_{j=0}^{n-4}\Bigl(U_{\mathrm{BF}}^{(j)} e^{-i\,\delta t\,J_{\mathrm{BF4}}\, Z_{j+2}} U_{\mathrm{BF}}^{(j)\dagger}\Bigr),
\end{align}
with $U_{\mathrm{BF}}^{(j)}=U_F^{(j+1)}U_B^{(j)}$, so that $V(1)=U_{\mathrm{B}}U_{\mathrm{A}}$ and $V(s)=\bigl(V(1)\bigr)^s$. Here $U_F^{(j)}$ is the local three-anyon recoupling unitary gate on window $(j, j+1, j+2)$, while $U_B^{(j)}$ is the corresponding braid unitary gate. Both are summarized in Appendix~\ref{app:fib-background}. 

The digital observable is the total Hamiltonian energy
\begin{equation}
E_{\mathrm{dig}}(n,s)=\mathrm{Tr}\!\left[\rho_n(s) H_n\right].
\label{eq:digital-observable}
\end{equation}
The noiseless simulations and (noisy) quantum-hardware experiments cover $n=5,7,9,11$ qubits and Trotter steps $s=1,\ldots,6$, for a total shot-budget range $N\in\{2000,4000,8000,16000\}$.

\subsection{Optimized-state VQE circuit}

The second benchmarking system we used is a quantum state whose optimal circuit parameters were obtained from a VQE noiseless simulation. This help isolates measurement quality from optimizer dynamics by fixing the variational parameters before the benchmark comparison. For each system size, we first prepare an optimized ``locked'' reference state with a hardware-efficient next-nearest-neighbour \texttt{Ry-Rz + CZ} ansatz and then benchmark \FusionRO\ and \PSQWC\ on that same state at matched shot budgets. This ansatz choice is consistent with the broader hardware-efficient VQE literature~\cite{kandala2017hardware, ayeni2025}. Starting from the same fusion-valid reference state $\ket{1\cdots 1}$, we write the ansatz state as
\begin{equation}
\ket{\psi_n(\theta)} = U_{\mathrm{NNN}}^{(d)}(\theta)\ket{1\cdots 1},
\label{eq:vqe-state}
\end{equation}
with depth $d=3$ for $n=5,7,9$ and $d=4$ for $n=11$. A convenient explicit form is
\begin{equation}
U_{\mathrm{NNN}}^{(d)}(\theta)
=
U_{\mathrm{rot,f}}(\theta^{(f)})
\prod_{k=1}^{d}
\left[
U_{\mathrm{ent}}^{(k)}
U_{\mathrm{rot}}^{(k)}(\theta^{(k)})
\right],
\label{eq:vqe-ansatz}
\end{equation}
where
\begin{align}
U_{\mathrm{rot,f}}(\theta^{(f)})
&= \prod_{j=0}^{n-1} R_z\!\left(\theta_{j,f}^{(z)}\right)R_y\!\left(\theta_{j,f}^{(y)}\right), \\
U_{\mathrm{rot}}^{(k)}(\theta^{(k)})
&= \prod_{j=0}^{n-1} R_z\!\left(\theta_{j,k}^{(z)}\right)R_y\!\left(\theta_{j,k}^{(y)}\right).
\end{align}
Each entangling block $U_{\mathrm{ent}}^{(k)}$ consists of one complete nearest-neighbour CZ sweep over even and odd bonds followed by a next-nearest-neighbour sweep $\prod_{j=0}^{n-3} \mathrm{CZ}_{j,j+2}$.
For each $n$, a separate noiseless exact-VQE stage minimizes the Hamiltonian expectation value over this ansatz, and the resulting best parameter vector is then frozen. The quantity estimated within each comparison cell is therefore
\begin{equation}
E_{\mathrm{VQE}}(n,\theta)=\langle \psi_n(\theta)|H_n|\psi_n(\theta)\rangle.
\label{eq:vqe-observable}
\end{equation}
The subsequent FR-versus-PS comparison therefore probes two measurement strategies on the same locked state rather than two different optimization histories.

\subsection{Matched shot budgets and hardware benchmark}

In both workflows the two methods are compared at the same total budget $N$. \FusionRO\ partitions that budget as evenly as possible across native local observables, while \PSQWC\ partitions the same budget as evenly as possible across its global QWC groups. The hardware benchmark uses the IBM superconducting quantum processor \texttt{ibm\_pittsburgh} as the compilation and execution target. No readout mitigation, dynamical decoupling, or resilience-level post-processing, and symmetry-verification-based error mitigation~\cite{BonetMonroig2020} is applied. The detailed hardware table is moved to Appendix~\ref{app:hardware-methodology}.

\section{Results: When Native Fusion Readout Helps and When It Fails}

The results we present are generated from the datasets obtained from both noiseless simulations and quantum-hardware experiments carried out on the IBM superconducting quantum processor \texttt{ibm\_pittsburgh}. In the noiseless case, each data cell uses $R=5$ replicates, i.e. independent realizations, whereas each hardware comparison cell uses $R=3$ replicates. The resulting hardware-side uncertainty estimates should therefore be seen as descriptive cellwise diagnostics rather than as asymptotic error bars.

We present results for the benchmark of the measurement methods executed over the workbench of digital Floquet time evolution and optmized VQE circuits.

\subsection{Digital benchmark: realized \MSE\ versus sampling variance}
\begin{figure*}[t]
\centering
\includegraphics[width=0.485\textwidth]{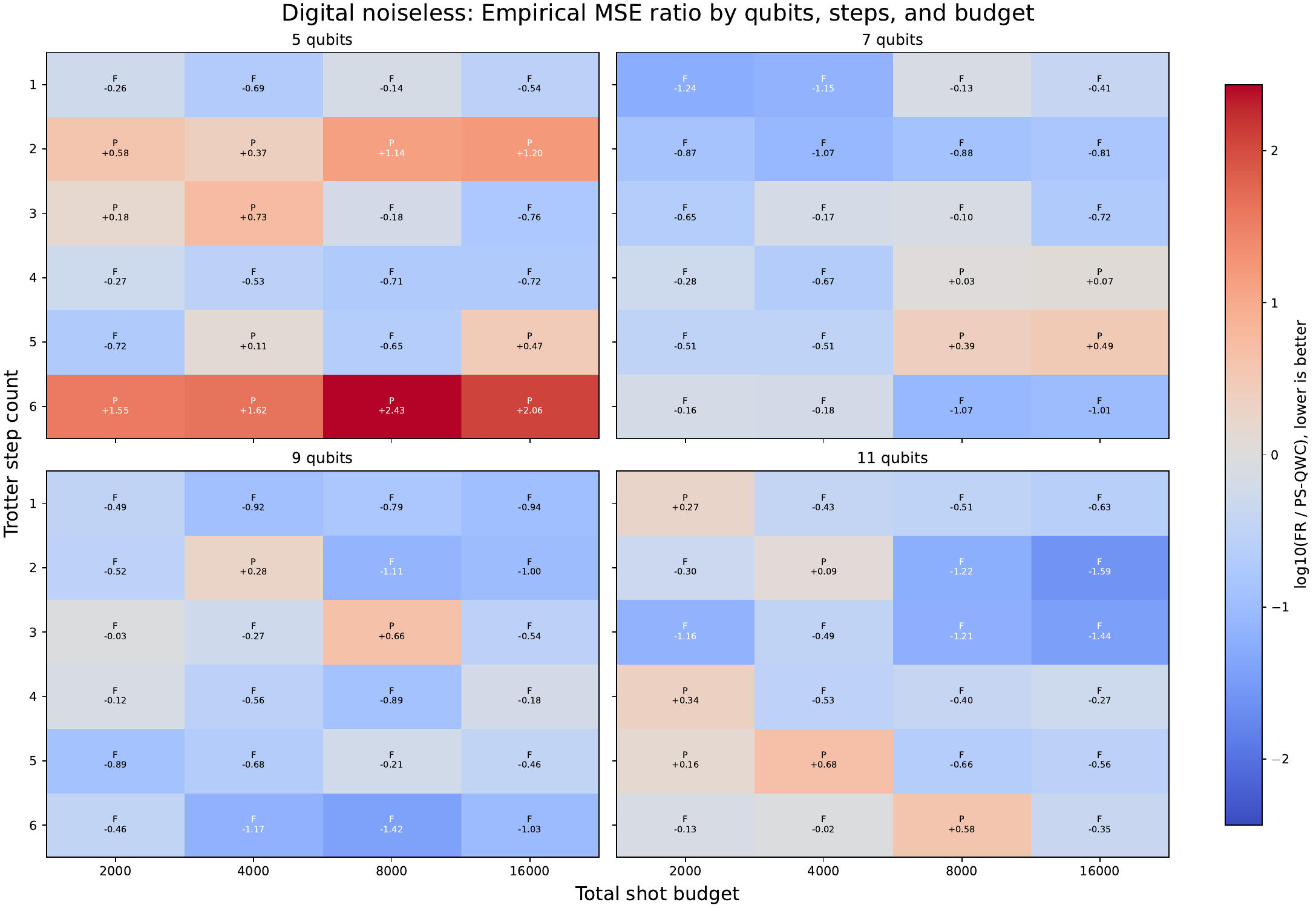}\hfill
\includegraphics[width=0.485\textwidth]{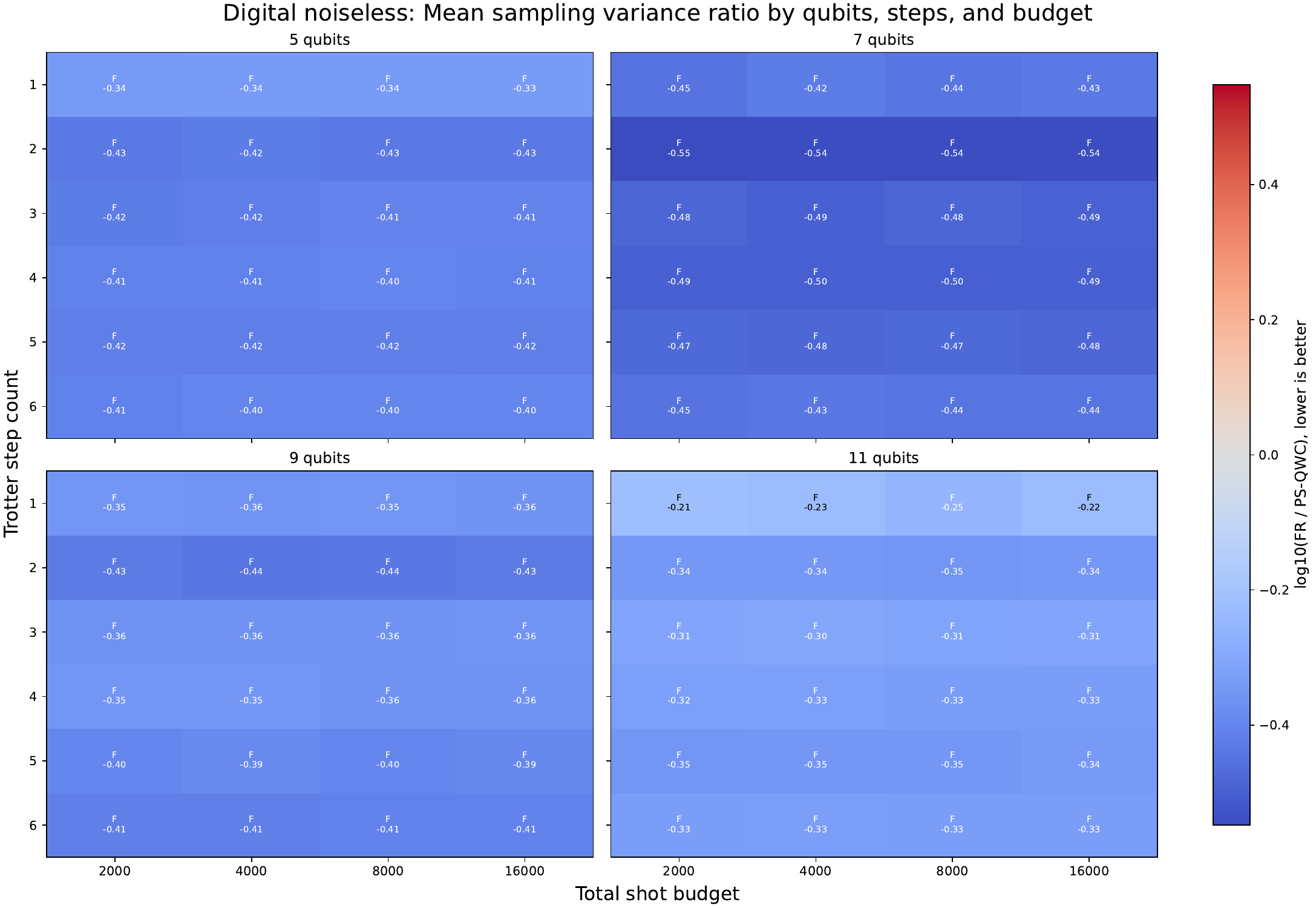}

\vspace{0.7em}

\includegraphics[width=0.485\textwidth]{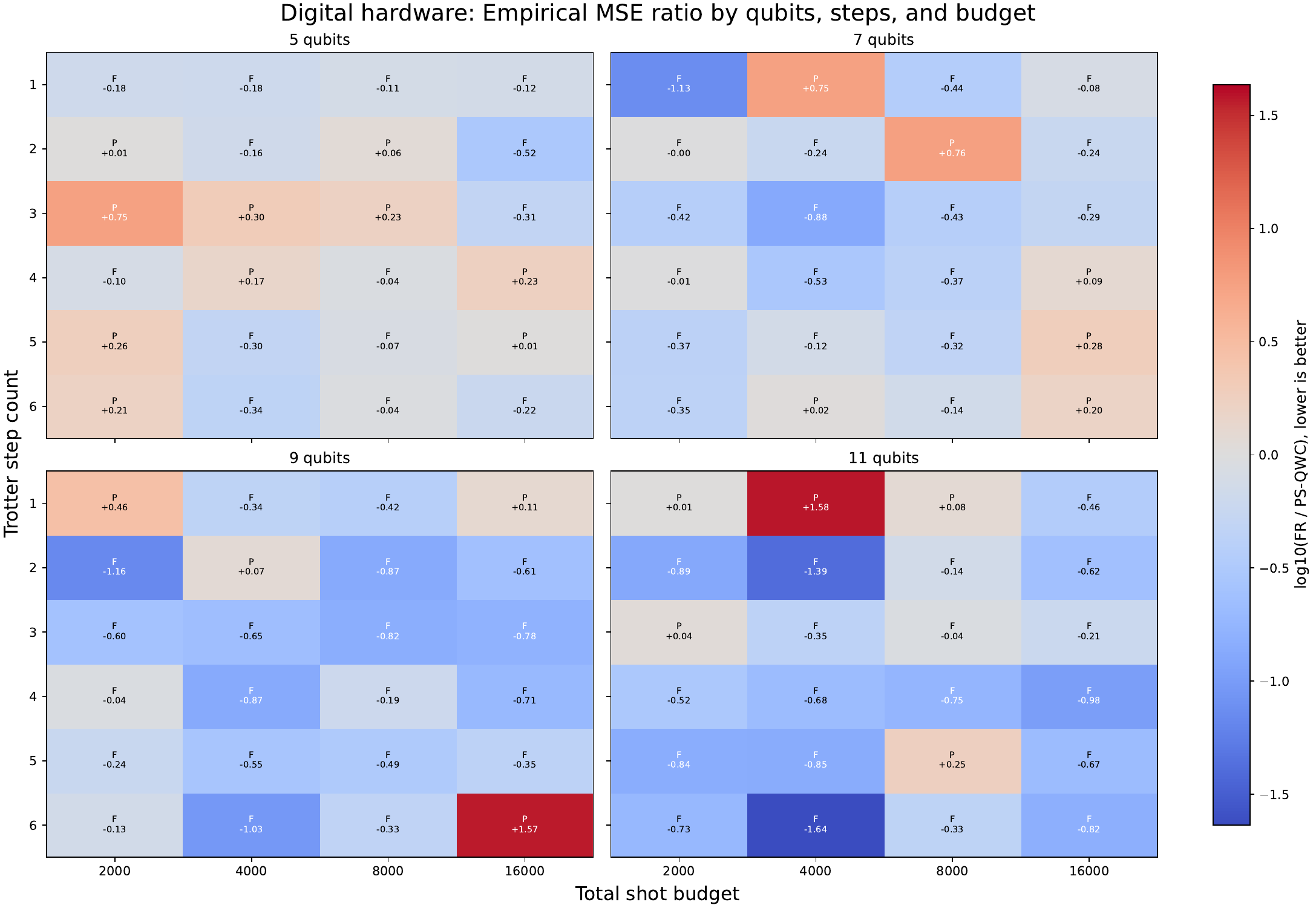}\hfill
\includegraphics[width=0.485\textwidth]{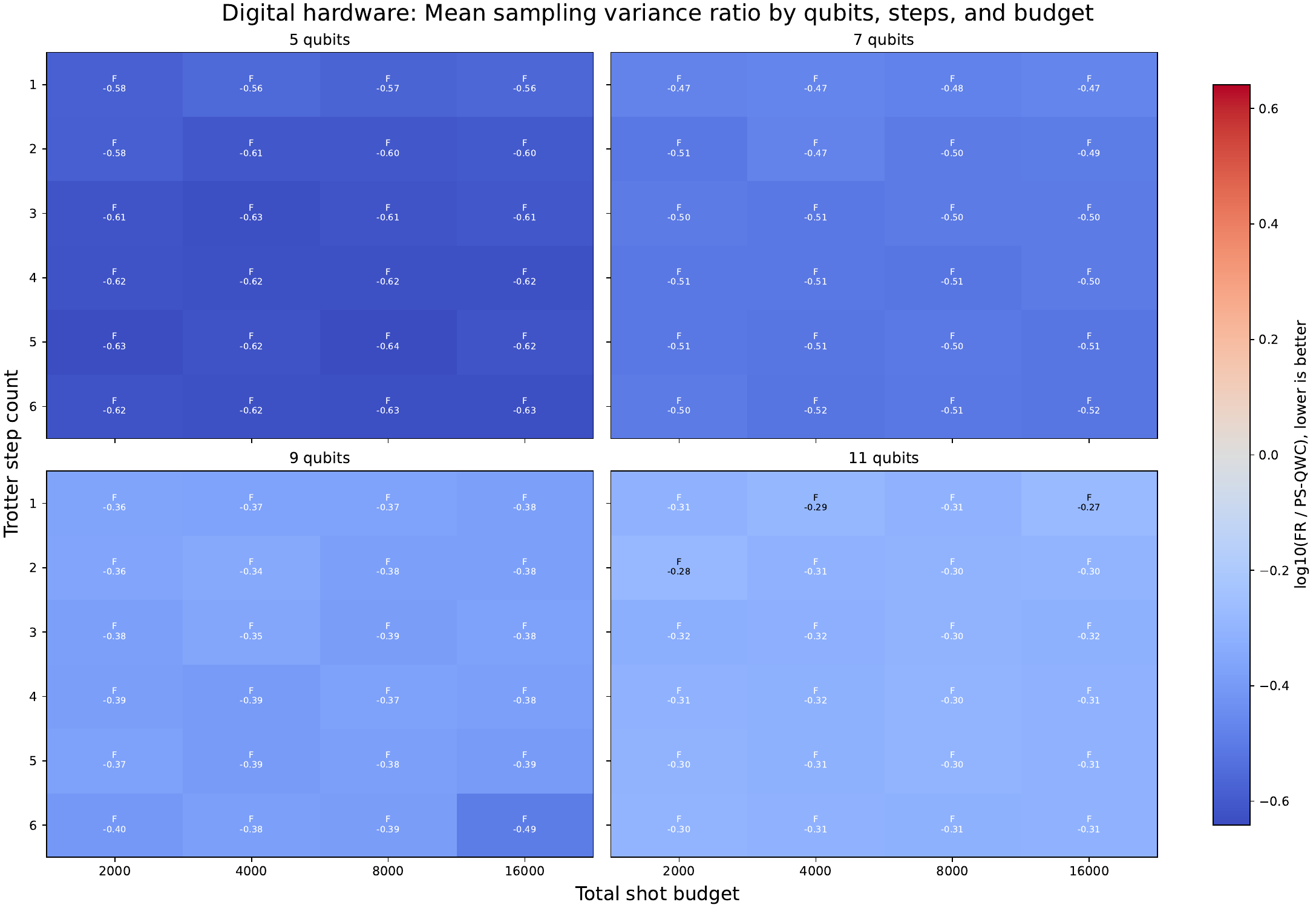}
\caption{Digital benchmark shown as realized estimator error versus covariance-aware sampling variance. Top row: noiseless empirical \MSE\ and mean sampling variance. Bottom row: hardware empirical \MSE\ and mean sampling variance. Lower values favor \FusionRO. The noiseless panels isolate the native sampling advantage; the hardware panels show where realized error partially ``washes out'' that advantage.}
\label{fig:digital-panels}
\end{figure*}

Figure~\ref{fig:digital-panels} presents the digital benchmark in the language of the estimator criterion. The left panels show realized empirical \MSE, while the right panels show the covariance-aware sampling variance from Eq.~\eqref{eq:sampling-var}. 

In the noiseless benchmark, \FusionRO\ wins 72 of 96 cells on empirical \MSE\ and all 96 cells on mean sampling variance. This is the cleanest evidence that the native measurement alignment lowers the covariance (yielding a higher precision) and usually lowers the realized estimator error (yielding a higher accuracy) when compiled hardware noise is absent.

On the hardware benchmark, the picture becomes more nuanced. \FusionRO\ still wins every cell on mean sampling variance, but it wins only 71 of 96 cells on empirical \MSE. That regime-level total is close to the noiseless value of 72 of 96, yet the underlying winner map is not: only 67 of 96 digital cells keep the same empirical-\MSE\ winner between the noiseless and hardware datasets, so 29 cells reverse winner once the compiled circuits are executed on hardware. The native sampling advantage therefore survives everywhere at the covariance level, while the realized estimator ordering becomes sensitive to hardware-induced changes in bias and empirical variance. The win count for \FusionRO\ is summarized in Table~\ref{tab:regime-summary}. 

The strongest PS-favorable tendency is concentrated in the $n=5$ sector rather than being distributed uniformly across qubit sizes. A plausible structural contributor is that in the digital Floquet circuit, each Trotter step performs ordered left-to-right sweeps over overlapping three-qubit F3 windows and four-qubit BF4 windows associated with the nearest-neighbour and next-neighbour interaction terms. At $n=5$ this leaves only five local terms per step, all boundary-touching, with the central qubit participating in every term, whereas the larger chains develop a genuine interior. The $n=5$ circuit is therefore structurally exceptional, which may help explain why the PS-favorable cells concentrate there.

With the present $R=3$ hardware realizations, many of the cellwise FR-versus-PS empirical-\MSE\ differences are small on the scale of their own replicate-level uncertainty. Across the 96 hardware cells, the median pooled standard error for $\Delta \widehat{\MSE}^{\mathrm{emp}}=\widehat{\MSE}^{\mathrm{emp}}_{\FusionRO}-\widehat{\MSE}^{\mathrm{emp}}_{\PSQWC}$ is $0.161$, while the median absolute gap $|\Delta \widehat{\MSE}^{\mathrm{emp}}|$ is only $0.114$. In other words, the typical winner margin is smaller than the uncertainty proxy attached to that margin, and only 28 of 96 cells have a gap larger than the pooled standard error. The close 71/96-versus-72/96 regime totals should therefore not be taken as evidence that the hardware and noiseless digital datasets behave identically; they reflect many weakly separated cells and a balance of cellwise reversals rather than an unchanged winner pattern.

\subsection{Optimized-state VQE benchmark: the sharpest hardware reversal}

In order not to confound the measurement benchmark test  with a classical optimizer routine inherent in every VQE workflow, we first obtain the optimal ground state from a classical VQE simulation of the Hamiltonian, and then used that optimal parameters value to reconstruct the optimized circuit on either a simulator or quantum hardware before feeding it to the measurement benchmark workflow. We call such a circuit either a ``locked'' or optimized state.

\begin{figure*}[t]
\centering
\includegraphics[width=0.88\textwidth]{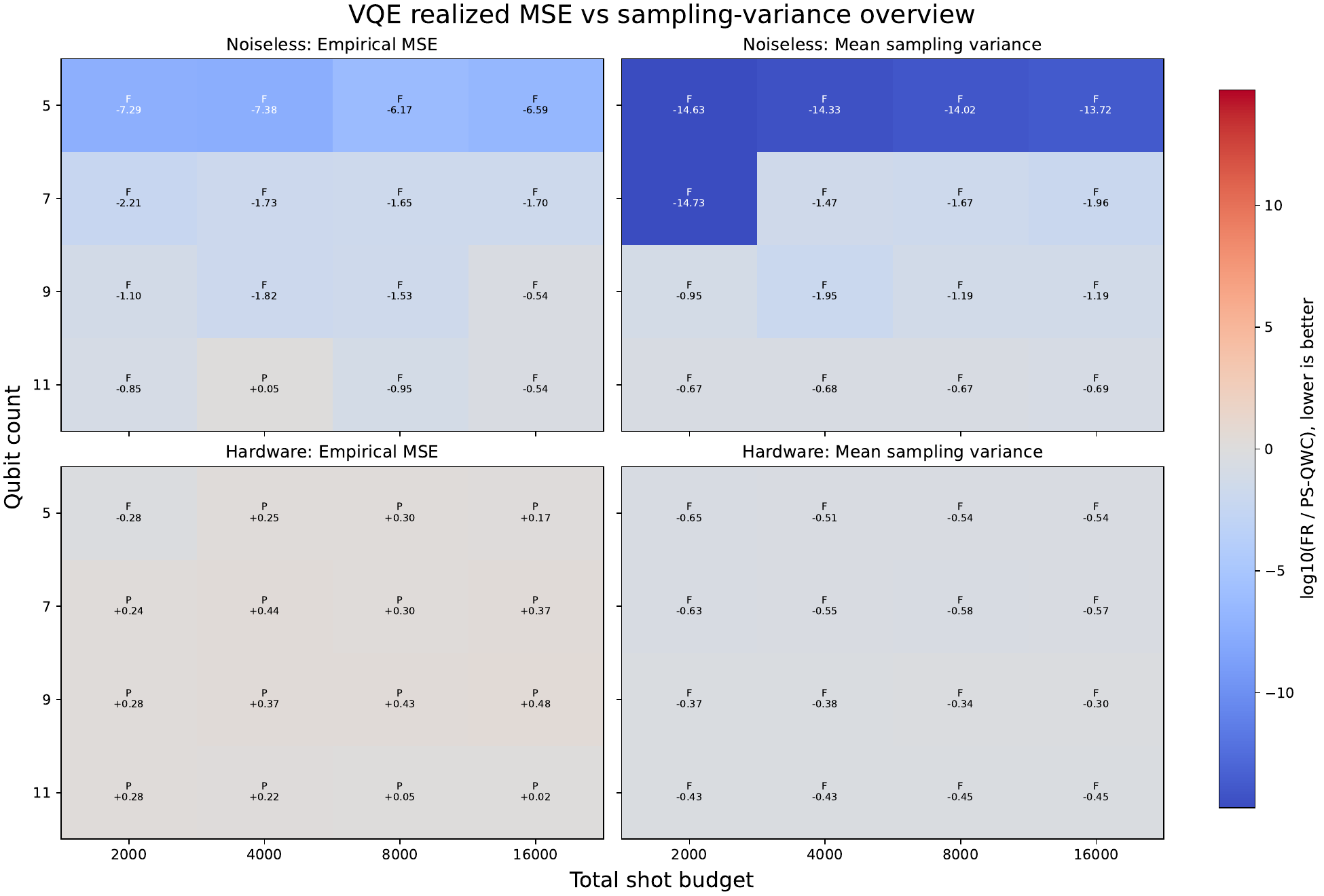}
\caption{Optimized-state VQE benchmark in the same realized-error-versus-sampling language used in the main text and also Figure~\ref{fig:digital-panels}. The noiseless comparison follows the native-measurement expectation, but the hardware comparison reverses on realized \MSE\ while leaving the covariance-aware sampling advantage of \FusionRO\ intact.}
\label{fig:vqe-panels}
\end{figure*}

The optimized-state VQE benchmark Figure~\ref{fig:vqe-panels} presents an entirely different view of the native-measurement method test. In the noiseless data, \FusionRO\ wins 15 of 16 cells on empirical \MSE\ and all 16 cells on mean sampling variance. The native readout logic therefore behaves as expected when the comparison is not polluted by compiled hardware error.

The hardware optimized-state VQE benchmark reverses that ordering on realized error. \PSQWC\ wins 15 of 16 cells on empirical \MSE, even though \FusionRO\ again wins all 16 cells on mean sampling variance. This is the clearest regime in the present benchmark where a genuine native sampling advantage survives at the covariance level but fails to determine the realized estimator error on hardware.

This optimized-state VQE reversal is also better resolved than the mixed digital hardware pattern. With the same $R=3$ hardware realizations, the median pooled standard error of the FR-minus-PS empirical-\MSE\ difference across the 16 cells is $0.917$, whereas the median absolute empirical-\MSE\ gap is $1.83$; 11 of 16 cells have $|\Delta \widehat{\MSE}^{\mathrm{emp}}|$ larger than that pooled error estimate. Table~\ref{tab:hardware-methodology} also points to the likely circuit-level reason: the variational circuit is shallow, so the appended FR measurement layer remains visibly larger than the PS-QWC layer. For instance, the minimum-to-maximum total two-qubit gate counts across the set of transpiled FR versus PS-QWC circuits at the n = 5 qubits are 98–222 versus 51–54 respectively, and 240–370 versus 190–193 at n = 11 qubits. The VQE reversal on the hardware could therefore be primarily coming from the cost of appending a larger FR measurement circuit to a shallow variational circuit, whereas the PS measurement circuit is smaller.

\begin{table}[t]
\caption{Regime-level summary of the benchmark. The reported numbers are the total of the cells in Figure~\ref{fig:digital-panels} where \FusionRO\ wins by \MSE\ and covariance-aware sampling variance. }
\label{tab:regime-summary}
\begin{ruledtabular}
\small
\begin{tabular}{lcc}
Benchmark & Empirical \MSE & Sampling var. \\
Digital, noiseless & 72/96 & 96/96 \\
Digital, hardware & 71/96 & 96/96 \\
optimized-state VQE, noiseless & 15/16 & 16/16 \\
optimized-state VQE, hardware & 1/16 & 16/16
\end{tabular}
\end{ruledtabular}
\end{table}

\subsection{Budget scaling values}

\begin{figure*}[t]
\centering
\includegraphics[width=0.95\textwidth]{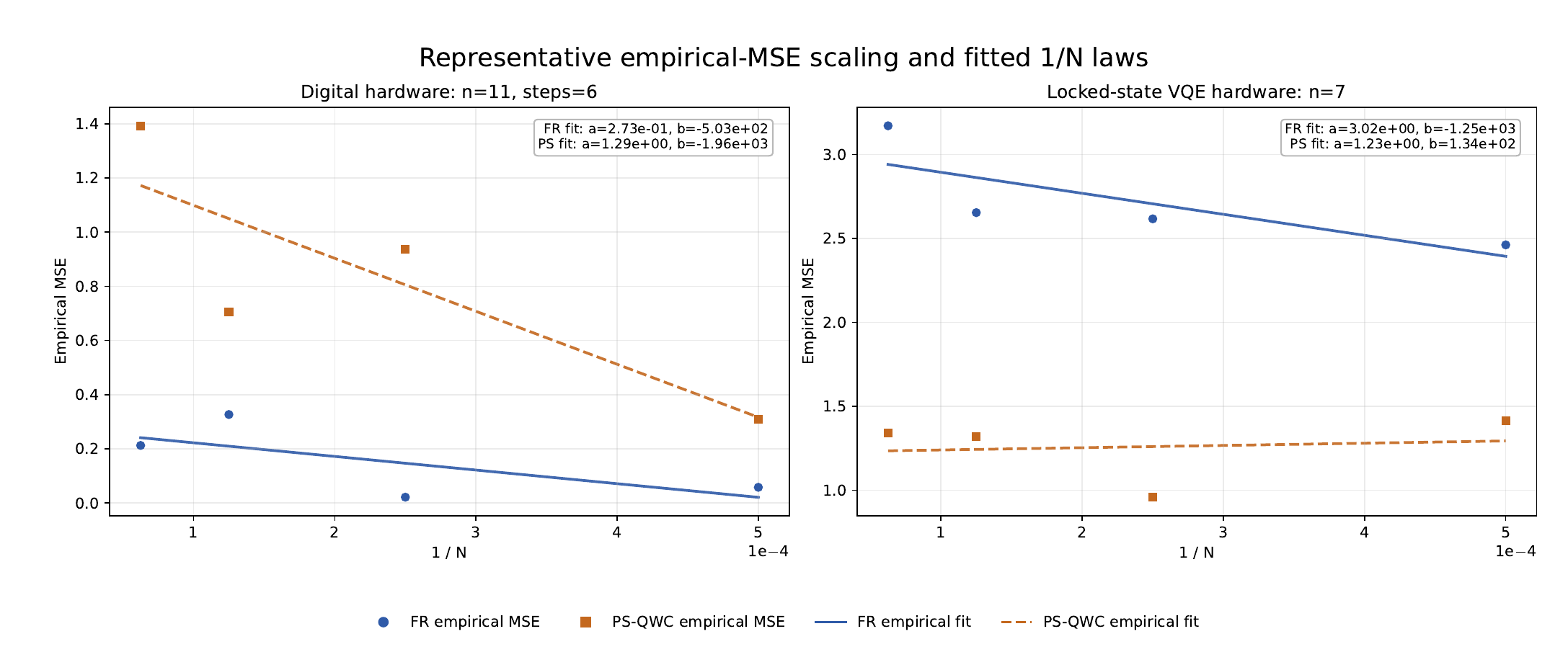}
\caption{Representative budget-scaling plots for one digital hardware cell and one optimized-state VQE hardware cell. Marker-only points show the empirical \MSE\ at the sampled budgets, using circles for \FusionRO\ and squares for \PSQWC. The overlaid solid FR and dashed PS-QWC lines are least-squares fits of the form $\alpha_m + \beta_m/N$ to the empirical mean-squared-error traces. The figure makes the fitted empirical-\MSE\ scaling visible directly in the budget traces.}
\label{fig:representative-scaling}
\end{figure*}

Heatmap summaries are efficient description of regime structure, but the budget dependence is easier to read directly from representative cells. Figure~\ref{fig:representative-scaling} therefore shows one digital hardware cell where \FusionRO\ remains favored and one optimized-state VQE hardware cell where the realized-error comparison flips toward \PSQWC. In both panels the horizontal variable is $1/N$, the empirical-\MSE\ values are shown as marker-only data points, and the fitted empirical-\MSE\ trends become ordinary line models. The digital representative cell is the hardware point $(n,s)=(11,6)$, where the highest-budget empirical difference remains negative in favor of \FusionRO. The optimized-state VQE representative cell is the $n=7$ hardware point, where the highest-budget empirical difference is positive in favor of \PSQWC\ while the corresponding sampling-variance difference remains negative in favor of \FusionRO. These plots make the central claim concrete: the native-measurement advantage lives in the covariance-sensitive part of the estimator, but the realized hardware comparison can still be controlled by the non-sampling contribution.

 Moreover, those plots reveal the scaling behaviour of the empirical MSE with the shot budget $N$, and its regime-dependence, which implies there might exist a crossover point between where either \FusionRO\ or \PSQWC\  lowers the empirical MSE. In the notation of Sec.~III, the overlaid FR-solid and PS-dashed fits are interpreted as empirical realizations of $\widehat{\MSE}_m(N)\approx \alpha_m + \beta_m/N$, with $\alpha_m$ capturing the effective fixed-budget error floor and $\beta_m$ capturing the covariance-sensitive slope together with hardware-sensitive $1/N$ contributions. Read this way, the digital representative cell is consistent with a favorable effective slope for \FusionRO\ that survives across the sampled budgets (i.e. sampling advantage), whereas the optimized-state VQE cell isolates the realized empirical-\MSE\ reversal directly. We can use the values obtained from the empirical fits to estimate the crossover point $N_c$ in the budget range (or outside of it), using the crossover criterion $N_c^{\mathrm{fit}}=|\Delta \beta|/\Delta \alpha$ [as explained in Sec.~\ref{Sec:Measurement criterion}, Eq.~\eqref{eq:crossover}] where the FR-minus-PS fit obeys $\Delta \alpha>0$ and $\Delta \beta<0$ (subsequently called the ``sign structure''). For the digital representative cell, the $N_c \approx 1.42\times 10^3$, which is below the budget range considered $N\in[2000,16000]$. For this case, \PSQWC\ favors shot budget $N$ less than $N_c$ while \FusionRO\ favors otherwise. For the representative optimized-state VQE, the $N_c \approx 7.73 \times 10^2$, and \FusionRO\ favors shot budget $N$ less than $N_c$ while \PSQWC\ favors otherwise. This representative example shows the regime behavior of MSE for depending on the type of the physical problem being considered. 
 
 %\rev{The fitted scaling also let us determine the crossover point $N_c$ from the crossover criterion in Eq.~\eqref{eq:crossover}. Using the empirical fits for each hardware scaling cell, we estimate a cellwise crossover budget from the fitted coefficient differences, $N_c^{\mathrm{fit}}=|\Delta \beta|/\Delta \alpha$ when the FR-minus-PS fit obeys $\Delta \alpha>0$ and $\Delta \beta<0$. Operationally, for one such scaling cell, $N_c^{\mathrm{fit}}$ is the fitted total-shot budget at which the FR and PS empirical-\MSE\ trend lines are equal: below that budget the fitted FR curve lies lower, while above it the fitted PS curve lies lower. 

We extended this scaling analysis to the entire dataset in our work. In the digital hardware data the sign structure appears in 4 of 24 scaling cells, and all four fitted crossovers lie inside the sampled budget range $N\in[2000,16000]$, with fitted values from $2.61\times 10^3$ to $8.57\times 10^3$ and median $4.53\times 10^3$. In the optimized-state VQE hardware data it appears in 3 of 4 scaling cells, and the median fitted value is $9.94\times 10^2$. The observed regime difference is therefore consistent with the same fitted criterion: in digital hardware the FR-favorable slope can still matter inside the sampled budgets, whereas in optimized-state VQE hardware the larger FR measurement circuit pushes the estimated crossover to very small budgets, making the MSE lean towards the PS-favored side of the reversal at high budgets. The crossovers are summarized in Table.~\ref{tab:crossover values}. These fitted crossover budgets represent descriptive thresholds in the present workload, compilation, and backend combination rather than as universal FR-versus-PS constants.

\begin{table}[t]
\caption{Regime crossover point $N_c$ obtained from the linear fit values for the crossover criterion Eq.~\ref{eq:crossover} for both digital Floquet and optimized-VQE state on hardware. }
\label{tab:crossover values}
%\begin{ruledtabular}
\small
\begin{tabular}{@{}lcc@{}}
\hline
Regime & Valid/total & Median $N_c$ [min, max] \\
\hline
Digital hw & 4/24 & $4.53\times10^{3}$ [$2.61\times10^{3}$, $8.57\times10^{3}$] \\
VQE hw & 3/4 & $9.94\times10^{2}$ [$7.73\times10^{2}$, $3.52\times10^{3}$] \\
\hline
\end{tabular}%
%\end{ruledtabular}
\end{table}

Appendix~\ref{app:selected-crossover-grids} records these fitted empirical-\MSE\ plots of selected digital hardware cells $(n,s)\in\{5,7,9,11\}\times\{1,3,6\}$ and the optimized-state VQE hardware cells $n=5,7,9,11$, with per-panel annotations indicating whether the fitted positive-budget crossing lies below, within, or above the sampled range.

Appendix~\ref{app:resource-proxy} analyzes the correlation between the method-dependent compiled measurement layer and the common error-ratio observable $\log_{10}(\mathrm{MSE}_{\mathrm{FR}}/\mathrm{MSE}_{\mathrm{PS}})$. In the digital hardware benchmark the budget-conditioned correlation is step dependent: for the three representative panels one finds $r=+0.14$ at $s=1$, $r=-0.45$ at $s=3$, and $r=-0.29$ at $s=6$. In the locked-state VQE hardware benchmark the corresponding budget-conditioned correlation is positive, $r=+0.46$. The count-versus-error relation is therefore weak and non-universal in the digital Floquet workload, but becomes positively aligned with the PS-QWC-favorable error ratio in the shallow VQE regime.

\section{Discussion}

The benchmark supports a practical measurement-design statement for compiled topological models. In the Fibonacci-chain setting, fusion readout is the physically natural rule and it systematically lowers the covariance-driven sampling term. The hardware results show, however, that this advantage is not by itself a reliable predictor of realized estimator quality once the required basis changes are compiled onto a noisy qubit backend.

That is why the comparison has to be organized around fixed-budget \MSE\ rather than covariance alone. If the question were only whether \FusionRO\ reduces the sampling term, the answer would be uniformly yes in the present data. The harder question is whether that benefit survives at fixed total budget after transpilation and hardware execution. Here the answer is regime dependent: the digital benchmark remains mostly favorable to \FusionRO, while the VQE-type circuit hardware benchmark shows a clear reversal toward \PSQWC.

The key factor that may explain the difference between the two workbench systems is that the variational state-preparation circuit is shallow, leaving the method-dependent measurement layer relative deep: at the representative VQE endpoints of Table~\ref{tab:hardware-methodology}, FR uses 98--222 two-qubit gates versus 51--54 for PS-QWC at $n=5$ and 240--370 versus 190--193 at $n=11$.\footnote{The count ranges because the logical gates implemented for the Fibonacci chain after being mapped to a sparse heavy-hex coupling graph do not transpile to the same 2Q count.} By contrast, the large digital endpoint already contains about $2.18\times10^4$ two-qubit gates for both methods, so the extra FR measurement layer is only a relatively small correction there. The optimized-state VQE reversal is therefore explained primarily by the larger depth of the FR measurement circuit relative to the smaller PS measurement circuit once both are appended to a shallow variational state-preparation circuit.

An important point worth noting in on the role of \PSQWC\ in the benchmark programme of this paper. It is an interpretable Pauli-frame reference baseline, not a claim about the globally best generic Pauli strategy. The point of the paper is instead to diagnose whether a measurement rule that is native to the target topological Hamiltonian remains advantageous after compilation to qubit hardware. This qualifier matters directly for the crossover scale $N_c$ in Eq.~\eqref{eq:crossover}. Stronger Pauli-side estimators, including shadow-style protocols or more aggressive grouping and ordering heuristics, could shift that scale by lowering the Pauli-side sampling or measurement overhead~\cite{huang2020predicting,Hadfield2022,verteletskyi2020measurement,gokhale2020optimization,huggins2021efficient,Crawford2021,Zhao2021,YenFullCommuting2023}. The present crossover boundaries should therefore be read as conditional on the interpretable \PSQWC\ baseline adopted here rather than as universal thresholds for every Pauli-side strategy.

The same scope applies to the hardware evidence itself. The research work is a descriptive benchmark built from a fixed workload suite, and finite replicate counts; it is not intended as high-power statistical inference about all compilers, all devices, or all Pauli-measurement baselines. Even so, it isolates a broader lesson relevant to topological simulation and topological engineering on qubit-native platforms: if one wants to measure projector- or fusion-based observables more natively, the basis changes must be co-designed with compilation, layout, and noise.

A related scaling point is worth stating explicitly. For the Fibonacci-chain observables studied here, each local fusion-readout basis change is logically a fixed-window primitive built from a constant number of $F$- and $R$-move blocks, so the ideal logical size of a single basis-change measurement does not itself grow extensively with $n$. What grows linearly with system size is the number of local Hamiltonian terms, and hence the number of distinct measurement circuits needed to reconstruct the full estimator. On sparse qubit hardware, routing and placement can then add further size-dependent compiled overhead on top of that local logical structure, so the practical compiled depth can increase more strongly than the ideal anyonic description alone would suggest.

\section{Conclusion}

The right benchmark quantity for native measurement in this setting is the fixed-budget mean-squared error of the full energy estimator, not covariance in isolation. On that metric, fusion readout delivers a genuine covariance-driven sampling advantage for Fibonacci-chain workloads and can substantially improve realized performance in noiseless and some hardware regimes. At the same time, the hardware benchmark shows that this advantage does not automatically survive compiled NISQ execution: basis-change overhead and hardware noise can erase or even reverse it.

The practical lesson is broader than this specific model. For topological and projector-dominated problems hosted on qubit-based processors~\cite{levin2005stringnet,Google2023,Quantinuum2024,XuFibonacci2024}, the key question is not only whether a native observable basis is physically natural, but whether it remains advantageous after compilation to the host hardware. The present results provide a concrete estimator-level workflow for answering that question.

From a longer-term fault-tolerant perspective, much of this tension is likely to be a NISQ-era effect. Once logical errors are pushed well below the target statistical uncertainty, extra basis-change depth should become more of a resource overhead than an accuracy limiter~\cite{fowler2012surface,campbell2017roads,litinski2019game}. That is why the natural next step is not to abandon native readout, but to make it more hardware-aware and to compare it against stronger Pauli estimators on a broader set of devices.

\begin{acknowledgments}
Thanks to colleagues at the Department of Physics at Maynooth University for helpful discussions and feedback during the early stages of this project. I am especially grateful to Joost Slingerland and Jiri Vala for their academic mentorship.

This work was supported in part by Enterprise Ireland through the DTIF programme of the Department of Business, Enterprise, and Innovation under project QCoIr: Quantum Computing in Ireland, A Software Platform for Multiple Qubit Technologies (No. DT 2019 0090B). In addition, I acknowledge the use of IBM Quantum Credits for this work. The views expressed are those of the author and do not reflect the official policy or position of IBM or the IBM Quantum team.
\end{acknowledgments}

\section*{Code Availability}
The public code release associated with this work is available as Ref.~\onlinecite{native_fusion_code_GitHub}.

\appendix

\section{Fibonacci anyon background and \texorpdfstring{$F$/$R$}{F/R} move conventions}
\label{app:fib-background}

This appendix fixes the minimal conventions needed to interpret the braid-plus-recouple local primitives used throughout the benchmark.

\subsection{Fibonacci \texorpdfstring{$F$/$R$}{F/R} data and the induced braid (B-move) operator}

We work with Fibonacci anyons with topological charge set $\{\mathbf{1},\tau\}$ and fusion rule~\cite{nayak2008nonabelian,pachos2012introduction,pfeifer2010simulation,singh2014matrix,ayeni2016simulation,kirchner2023numerical}
\begin{equation}
\tau \times \tau = \mathbf{1} + \tau,
\end{equation}
with quantum dimension $d_{\tau}=\varphi$ where $\varphi=(1+\sqrt{5})/2$ is the golden ratio.

The nontrivial recoupling move is the $F$ move for three $\tau$ anyons with total charge $\tau$. In the standard basis where the intermediate fusion channel is $x\in\{\mathbf{1},\tau\}$, the corresponding $2\times 2$ matrix is
\begin{equation}
F \equiv F^{\tau\tau\tau}_{\tau}
=
\begin{pmatrix}
\varphi^{-1} & \varphi^{-1/2} \\
\varphi^{-1/2} & -\varphi^{-1}
\end{pmatrix}.
\label{eq:fib-F-matrix}
\end{equation}
Braiding two $\tau$ anyons acts diagonally in the direct-fusion basis, with phases determined by the fusion outcome. The corresponding $R$ move can be represented by
\begin{equation}
R \equiv R^{\tau\tau}
=
\begin{pmatrix}
e^{-4\pi i/5} & 0 \\
0 & e^{3\pi i/5}
\end{pmatrix}.
\label{eq:fib-R-matrix}
\end{equation}

In the fusion-tree basis used for the chain, an elementary braid in the non-direct basis is represented by conjugating the direct-fusion $R$ move into the chosen association of the fusion tree. For the local three-anyon sector this gives the $2\times 2$ braid matrix
\begin{equation}
B \equiv F^{\dagger} R F.
\label{eq:fib-B-matrix}
\end{equation}
The many-body operators $F_i$ and $B_i$ in Eq.~\eqref{eq:local-terms} are the corresponding embeddings of these local primitives into the appropriate windows of the chain.

\paragraph{Circuit realization (summary).}
In our qubit fusion-path encoding, these moves are implemented as small fixed circuits acting on short windows: the $R$ move is a diagonal phase gate in the direct-fusion frame, the $F$ move is the controlled implementation of Eq.~\eqref{eq:fib-F-matrix}, and the braid circuit is the composition
\begin{equation}
U_B = U_F\,\bigl(\mathbb{I}\otimes U_R\otimes \mathbb{I}\bigr)\,U_F^{\dagger}.
\label{eq:B-circuit-identity}
\end{equation}
For Fibonacci data the $2\times2$ recoupling matrix in Eq.~\eqref{eq:fib-F-matrix} is real, symmetric, and involutory, so $U_F^{\dagger}$ may be implemented by the same local circuit as $U_F$. These are the implementation-level identities behind the basis changes $U_i=F_{i+1}B_i$ in Eq.~\eqref{eq:bf4-traces} and $U_{\mathrm{BF}}^{(j)}=U_F^{(j+1)}U_B^{(j)}$ in the product-formula expression.

\subsection{Worked example: a braid-plus-recoupling fusion-projector term (BF4)}

The simplest local operators in Fibonacci chains are fusion-channel projectors on neighboring anyons. In a fusion-tree basis these are naturally measured in a direct-fusion frame: an $F$ move recouples the fusion tree so that the relevant fusion outcome is represented as a single intermediate label, and the corresponding direct-fusion observable is a diagonal projector, or up to an affine shift a Pauli-$Z$ on the associated encoded qubit.

The BF4 benchmark term is a slightly more involved example in which the physical interaction is most naturally described after braiding and then recoupling into the direct-fusion frame. For a representative prepared state $\rho$, the operator from Eq.~\eqref{eq:bf4-conjugation} may be read as a next-nearest-neighbor fusion-projector term built from a local braid $B_i$ on the left pair together with the recoupling $F_{i+1}$ that brings the interacting pair into a direct-fusion configuration. In that frame, the observable $h_{i+1}^{\mathrm{DF}}$ is simple local projector-like.

For an arbitrary prepared state $\rho$, the corresponding local energy contribution admits the two equivalent trace forms
\begin{align}
\langle H_i^{\mathrm{BF4}}\rangle_{\rho}
&= \mathrm{Tr}\!\left(B_i^{\dagger}F_{i+1}^{\dagger} h_{i+1}^{\mathrm{DF}} F_{i+1} B_i\,\rho\right) \\
&= \mathrm{Tr}\!\left(h_{i+1}^{\mathrm{DF}}\,F_{i+1} B_i\,\rho\,B_i^{\dagger} F_{i+1}^{\dagger}\right).
\label{eq:bf4-appendix-traces}
\end{align}
These two forms make the measurement logic transparent. In \PSQWC, one measures a grouped Pauli decomposition of the mapped operator content entering Eq.~\eqref{eq:bf4-conjugation}. In \FusionRO, one instead appends the basis change $U_i = F_{i+1}B_i$ to the state-preparation circuit and measures the simple direct-fusion observable $h_{i+1}^{\mathrm{DF}}$ directly. This local worked example is the basic reason that fusion readout is physically natural for the Hamiltonian terms studied in the main text.

\section{Selected Crossover Scaling Grids}
\label{app:selected-crossover-grids}

\begin{figure*}[t]
\centering
\includegraphics[width=0.97\textwidth]{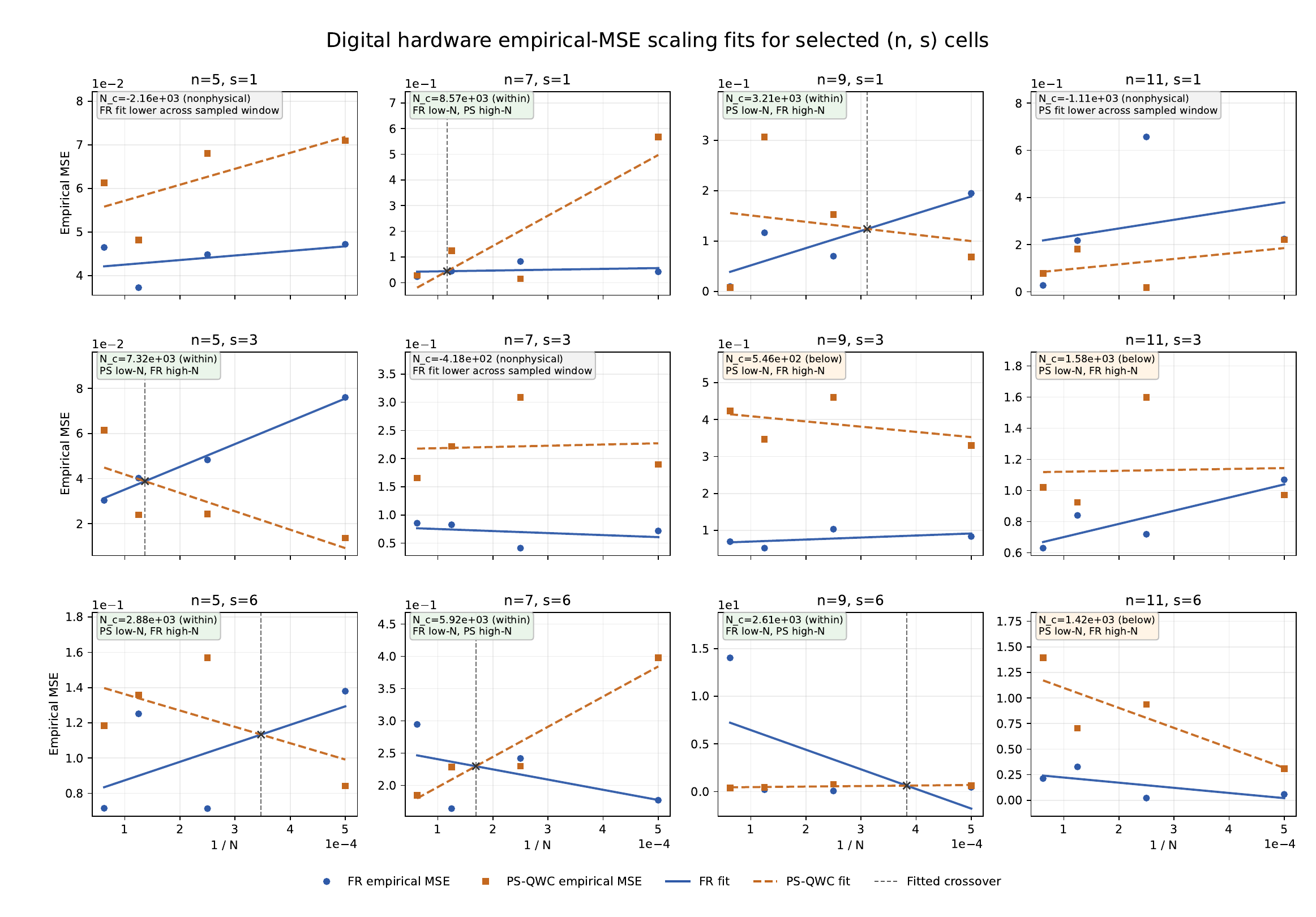}
\caption{Digital hardware empirical-\MSE\ scaling fits for the selected cells $(n,s)\in\{5,7,9,11\}\times\{1,3,6\}$. The point markers show the empirical \MSE\ values for \FusionRO\ and \PSQWC, while the solid and dashed lines  show the fitted laws $\alpha_m + \beta_m/N$, and dashed vertical lines mark fit positive-budget crossings that fall inside the sampled budget window. Panel annotations report the fitted crossing budget when it exists and indicate whether it lies below, within, or above the sampled range.}
\label{fig:appendix-digital-crossover-grid}
\end{figure*}

\begin{figure*}[t]
\centering
\includegraphics[width=0.97\textwidth]{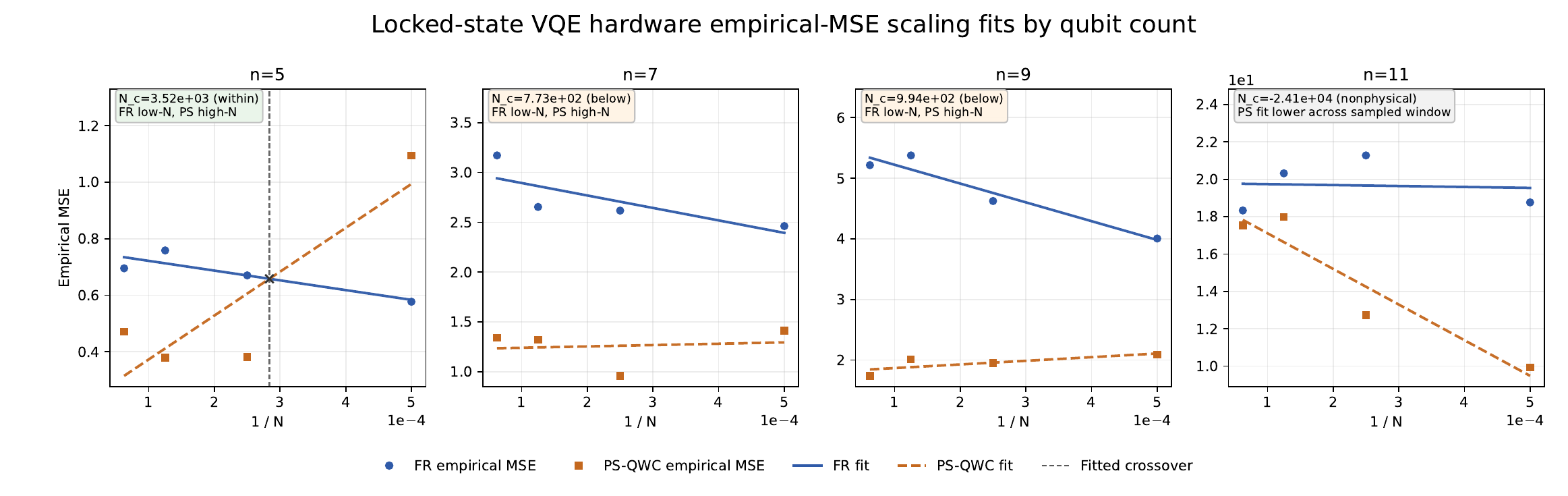}
\caption{Optimized-state VQE hardware empirical-\MSE\ scaling fits for qubit counts $n=5,7,9,11$. Point markers show the empirical \MSE\ values for \FusionRO\ and \PSQWC, while solid and dashed lines show the fitted laws $\alpha_m + \beta_m/N$, and dashed vertical lines mark fit positive-budget crossings that fall inside the sampled budget window. The VQE panels make clear that most fitted crossings for this workload already sit below the sampled budget range, consistent with the small fitted $N_c$ values reported in the main text.}
\label{fig:appendix-vqe-crossover-grid}
\end{figure*}

Figures~\ref{fig:appendix-digital-crossover-grid} and \ref{fig:appendix-vqe-crossover-grid} collect the fitted empirical-\MSE\ scaling plots for the selected hardware cells discussed in the crossover analysis. For the digital benchmark we show plots for the shallow, intermediate, and deep Trotter steps $s=1,3,6$ across qubit counts $n=5,7,9,11$. For the optimized-state VQE hardware benchmark, we show the analogous fits at $n=5,7,9,11$. In every panel the empirical \MSE\ data are indicated with point markers, while the least-squares fits of the form $\alpha_m + \beta_m/N$ are indicated with both solid lines (for \FusionRO) and dashed-solid lines (for \PSQWC), and the dashed vertical guide marks the fitted positive-budget crossing when it lies inside the sampled window. Panels whose fitted crossing lies below or above the sampled range are labeled accordingly, as well as panels with no positive fitted crossing. 

\section{Paired Measurement-Count Correlations}
\label{app:resource-proxy}

To characterize how compiled measurement overhead correlates with realized estimator performance, we analyze paired FR-minus-PS observables within each hardware cell. For a fixed workload, the state-preparation circuit is common to both measurement schemes, so the FR-minus-PS count difference removes the shared state-preparation contribution and isolates the method-dependent compiled measurement layer. We then correlate this paired count observable with the corresponding paired error observable while conditioning on the workload parameter that sets the dominant background scale: Trotter step $s$ in the digital Floquet benchmark and total shot budget $N$ in the locked-state VQE benchmark.

\begin{figure*}[t]
\centering
\includegraphics[width=0.94\textwidth]{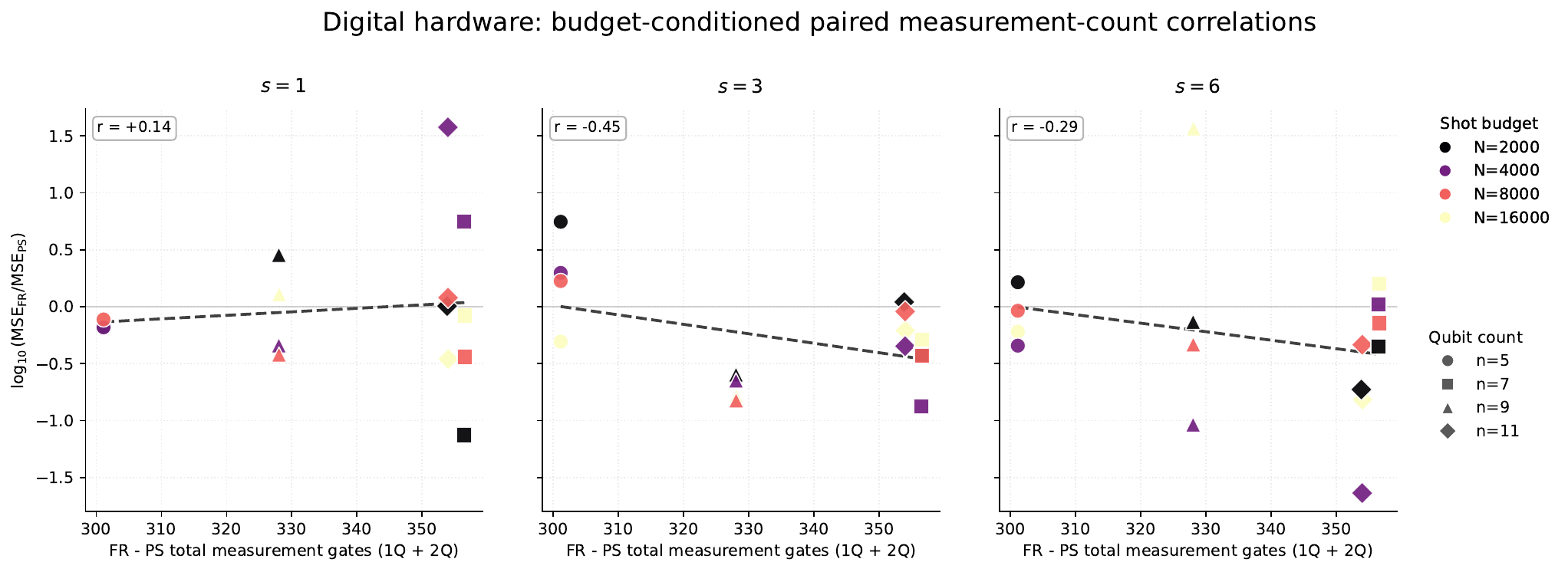}
\caption{Digital paired measurement-count correlations for the hardware Floquet benchmark. The three panels correspond to fixed Trotter steps $s=1,3,6$. Within each panel, each point is one fixed $(n,N)$ hardware cell. The horizontal axis is the FR-minus-PS total compiled measurement-gate count (1Q+2Q), and the vertical axis is $\log_{10}(\mathrm{MSE}_{\mathrm{FR}}/\mathrm{MSE}_{\mathrm{PS}})$ at the corresponding budget. Point color denotes shot budget, marker shape denotes qubit count, the dashed line is the budget-conditioned least-squares trend guide, and the inset reports the panelwise correlation coefficient $r$.}
\label{fig:digital-paired-measurement-correlation}
\end{figure*}

Figure~\ref{fig:digital-paired-measurement-correlation} resolves the digital hardware data into three fixed-step panels, $s=1,3,6$. Within each panel, every point is a fixed $(n,N)$ hardware cell and the ordinate is the instantaneous error-ratio observable $\log_{10}(\mathrm{MSE}_{\mathrm{FR}}/\mathrm{MSE}_{\mathrm{PS}})$ at that budget. The panelwise correlations are obtained after conditioning on shot budget, so each reported $r$ measures how the FR-minus-PS compiled measurement count varies with the FR-to-PS error ratio across qubit size at fixed $N$. For the total compiled measurement count (1Q+2Q), the resulting correlations are weak and step dependent: $r=+0.14$ at $s=1$, $r=-0.45$ at $s=3$, and $r=-0.29$ at $s=6$. Since negative values of the log-ratio correspond to FR-favorable performance, those 3-panel plots show that \FusionRO\ become more favorable with increased circuit depth at higher Trotter step value.

\begin{figure*}[t]
\centering
\includegraphics[width=0.94\textwidth]{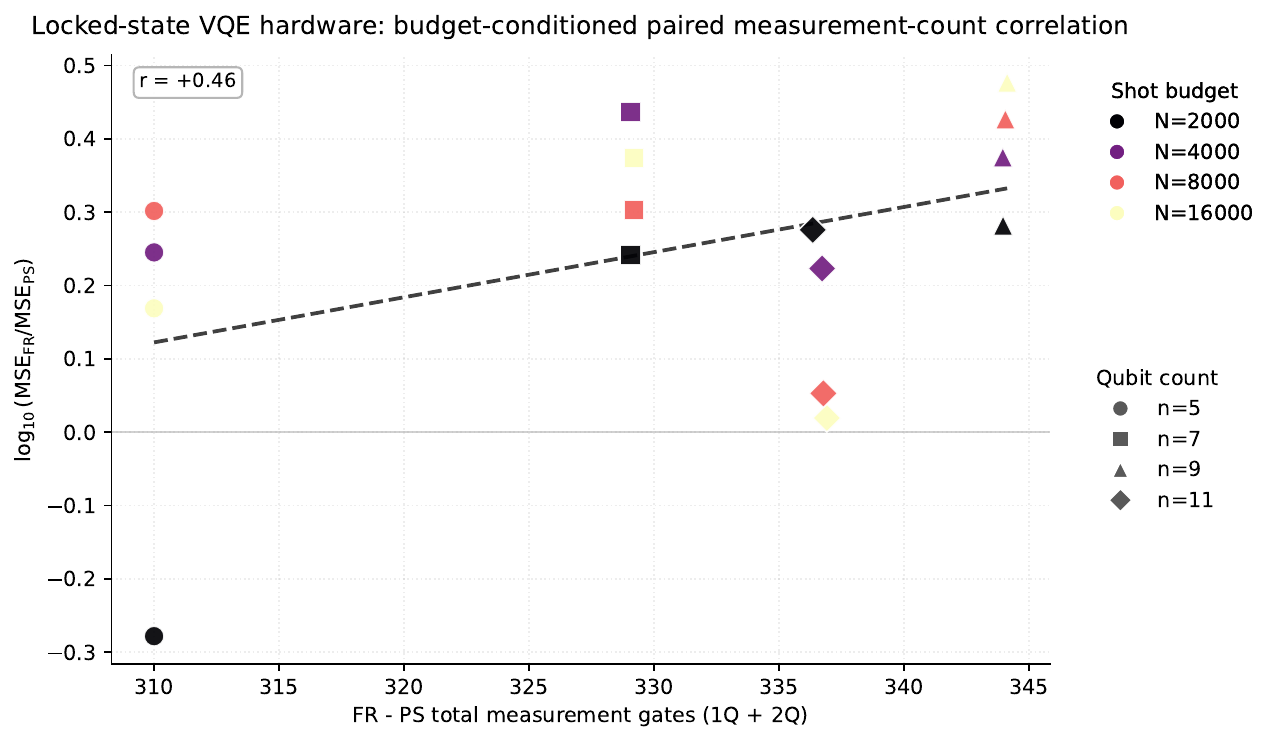}
\caption{Locked-state VQE paired measurement-count correlation for the hardware benchmark. Each point is one fixed $(n,N)$ hardware cell. The horizontal axis is the FR-minus-PS total compiled measurement-gate count (1Q+2Q), and the vertical axis is $\log_{10}(\mathrm{MSE}_{\mathrm{FR}}/\mathrm{MSE}_{\mathrm{PS}})$ at that budget. Point color denotes shot budget, marker shape denotes qubit count, the dashed line is the budget-conditioned least-squares trend guide, and the inset reports the correlation coefficient $r$.}
\label{fig:vqe-paired-measurement-correlation}
\end{figure*}

Figure~\ref{fig:vqe-paired-measurement-correlation} applies the same vertical observable to the locked-state VQE hardware data. Here each point is again a fixed $(n,N)$ cell, and the budget-conditioned correlation between the FR-minus-PS total compiled measurement count and $\log_{10}(\mathrm{MSE}_{\mathrm{FR}}/\mathrm{MSE}_{\mathrm{PS}})$ is positive, $r=+0.46$. The positive sign means that a larger compiled FR measurement layer tends to coincide with a larger PS-QWC-favorable error ratio, consistent with the shallow locked-state circuits leaving the method-dependent measurement block comparatively exposed.

The correlation structure therefore differs between the two hardware workloads. In the digital Floquet benchmark, the relation between compiled measurement-count difference and the FR/PS error ratio is weak but leans more to be negative at larger Trotter steps, indicating a weak correlation between the large FR measurement circuits and the observed error. In the locked-state VQE benchmark, the same paired observable is positively correlated with the error ratio on the common log-scale, indicating that compiled measurement overhead is a more relevant control parameter in the shallow-circuit regime.

Beyond the gate-count diagnostic itself, other plausible contributors to the hardware reversal include layout-sensitive routing on the heavy-hex graph, and the nonuniform assignment-error landscape across candidate qubits.

\section{Hardware Transpilation Methodology}
\label{app:hardware-methodology}

The hardware benchmark is descriptive rather than inferential due to the current prohibitive cost of running experiments on quantum hardware, but the compilation context still matters because the central question of the paper is whether the additional basis-change depth required by native readout survives under current-device constraints. Table~\ref{tab:hardware-methodology} collects the configuration of the hardware. 

That compiled-hardware perspective is also the one now relevant to recent processor-level demonstrations of non-Abelian braiding and topological-order engineering, where native logical observables must still be accessed through the compilation and noise profile of the host platform~\cite{Google2023,Quantinuum2024,XuFibonacci2024}.
All logical-circuit construction, transpilation, and execution were carried out in a Qiskit-based workflow~\cite{Qiskit}.

\begin{table*}[t]
\centering
\scriptsize
\setlength{\tabcolsep}{4pt}
\caption{Hardware-methodology details for the benchmark. The upper panel lists the \texttt{ibm\_pittsburgh} calibration. The lower panel reports the matched-budget measurement allocations and endpoint transpiled circuit-resource ranges for representative digital and optimized-state VQE workloads.}
\label{tab:hardware-methodology}
\begin{tabular}{|p{0.24\textwidth}|p{0.70\textwidth}|}
\hline
\textbf{Backend quantity} & \textbf{Value} \\
\hline
Backend and calibration date & \texttt{ibm\_pittsburgh}, calibration record 2026-05-04 12:56 UTC \\
\hline
Execution mode & IBM Quantum hardware campaign on \texttt{ibm\_pittsburgh}; queue delay is not part of the estimator analysis \\
\hline
Backend size & 156 qubits, 352 directed couplers \\
\hline
Basis gates & \texttt{cz,id,rx,rz,rzz,sx,x} \\
\hline
Transpilation and mitigation & Qiskit preset pass manager, optimization level 1; no readout mitigation, dynamical decoupling, or resilience-level post-processing \\
\hline
Readout assignment error & Mean 0.76\%, range 0.15\%--13.27\% across qubits \\
\hline
Native two-qubit calibration & \texttt{cz} gate; mean error 0.58\%, median 0.145\%, range 0.077\%--13.72\% across calibrated couplers \\
\hline
Excluded coupler note & 10 couplers reported placeholder \texttt{gate\_error}=1.0 entries in the backend calibration record and were excluded from the quoted \texttt{cz} mean. \\
\hline
\end{tabular}

\vspace{0.5em}

\resizebox{\textwidth}{!}{%
\begin{tabular}{|l|l|l|r|r|r|r|}
\hline
\textbf{Endpoint} & \textbf{FR allocation at $B=2000$} & \textbf{PS-QWC allocation at $B=2000$} & \textbf{FR depth} & \textbf{FR 2Q} & \textbf{PS depth} & \textbf{PS 2Q} \\
\hline
Digital $n=5$, $s=1$ & 5 observables, 400 each & 15 groups, 133--134 each & 2444--2711 & 1073--1200 & 2330--2363 & 1029--1053 \\
\hline
Digital $n=11$, $s=6$ & 17 observables, 117--118 each & 22 groups, 90--91 each & 30262--33745 & 21779--22125 & 30137--34553 & 21834--22059 \\
\hline
VQE $n=5$, $d=3$ & 5 observables, 400 each & 15 groups, 133--134 each & 189--469 & 98--222 & 84--94 & 51--54 \\
\hline
VQE $n=11$, $d=4$ & 17 observables, 117--118 each & 22 groups, 90--91 each & 223--523 & 240--370 & 127--188 & 190--193 \\
\hline
\end{tabular}%
}
\end{table*}

\clearpage
\bibliography{References}

\end{document}